# The role of the $C_2$ gas in the emergence of $C_{60}$ from the condensing carbon vapour

Shoaib Ahmad

*Government College University (GCU), CASP, Church Road, 54000 Lahore, Pakistan*

Kashif Yaqub and Afshan Tasneem

*Pakistan Institute of Nuclear Science and Technology (PINSTECH), P.O. Nilore, Islamabad,*

*Pakistan*

**ABSTRACT**

A model has been developed that illustrates $C_{60}$'s emergence from the condensing carbon vapour. It is shown to depend upon (i) the decreasing heats of formation for larger cages, (ii) exponentially increasing number of isomers for fullerenes that are larger than $C_{60}$, (iii) large cages' buckling induced by the pentagon-related protrusions that initiate fragmentation, (iv) the structural instability-induced fragmentation that shrinks large cages via $C_x \rightarrow C_{x-2} + C_2$ and (v) an evolving gas of $C_2$ that is crucial to the whole process. The model describes a mechanism for the provision and presence of plenty of $C_2$s during the formation and fragmentation processes. Fullerenes portrayed as 3D rotors have partition functions describing ensemble's entropy as a function of the fragmentation sequence. The bottom-up formations of large cages followed by the top-down cage shrinkage are shown to be stable, dynamical processes that lead to the $C_{60}$ dominated fullerene ensemble.

## INTRODUCTION

Various mechanisms for the formation of $C_{60}$ have been actively proposed and debated ever since its discovery. Theoretical and simulation studies have made good progress[1-10] but an all-encompassing model is still desirable that may illustrate the routes of the self organizing carbon clusters that eventually lead to the Buckyball and at the same time, describe the statistical mechanics of the associated negative entropy. Fullerenes-the closed cages of carbon have been shown to exist in various sizes and geometrical shapes. 28 fullerene point groups can describe all shapes and topographies of the closed cages[11]. Icosahedral group $I_h$ to which one isomer each of $C_{20}$, $C_{60}$, $C_{80}$, $C_{180}$, $C_{240}$, $C_{540}$… belong, has the highest symmetry. Do the perfectly spherical $C_{60}$ emerges "out of the chaos of carbon vapour" as Smalley suggested[12] or the dynamics of the condensing carbon vapour produce an ensemble of large closed cages that self organize by shrinking? Simulations of the shrinking large cages have demonstrated that this is a plausible route for the formation of $C_{60}$. Using statistical mechanical arguments, we show that the ensuing gas of $C_2$ provides the essential missing link for the grand canonical ensemble of the newly formed cages to deliver $C_{60}$ as the end product.

## DYNAMICS OF THE BOTTOM-UP CAGE FORMATION

Carbon vapor condenses with higher probabilities for clusters $C_x$ with increasing number of carbon atoms-$x$. Linear chains, rings and sheets mostly have single isomer for each type until $x \geq 20$ when closed cages start to form and thereafter, have an increasing number of isomers for each $x$. The cluster growth dynamics of the initial, bottom-up phase is dominated by large closed cages whose respective heats of formation $E_x$ reduce with the increasing $x$. The requirement for the cage's formation energy to reduce for the larger sizes ensures higher densities of the closed cages with ever increasing diameters. A cage's binding energy and the respective heat of formation and its formation is proportional to $\exp(-E_x/kT)$. A cluster forming carbon vapour environment was modeled by using MNDO technique[13] that used the $C_2$ addition route to the formation of cages. Fig. 1 shows three possible mechanisms for $C_{60}$ cage to be formed. The formation energy per carbon atom $E_x$ is plotted for all the cages between $x = 24$ and 60. Results from the open cage route versus the closed cage road are shown in the figure. The larger and closed cage clusters have lower formation energies and the trend continues for the higher fullerenes. The inset contains the results that are expected beyond $C_{60}$. This indicates higher probabilities for larger fullerenes to form in a condensing carbon vapor. One envisages a primary ensemble that has higher proportion of the larger fullerenes. How do these transform into their smaller and eventually the perfect spheroidal neighbor, is explored in later sections.

Fullerenes as a class of closed cages composed of twelve pentagons and varying number of hexagons have more than one way of arranging these on the spheroid's surface leading to many isomers per fullerene. The only exception is $C_{20}$ that has no hexagon and hence the twelve pentagons assemble as a dodecahedron. Larger fullerenes have increasing number of isomers $I_x$ for each $C_x$. Atlas of Fullerenes[11] provides a compilation of isomers for each fullerene. The isomeric distribution of fullerenes as a function of $x$ indicates an increasing exponential dependence as shown in Fig. 2. We have chosen the range $x = 60$ to 100 to illustrate the $I_x$ versus $x$ that will be used to develop the model for fullerene growth. An exponential distribution is fitted to the isomeric data that yields $I_x \sim \exp(l_o x)$, where $l_o$ along with $x$, characterizes the dynamics of the distribution of fullerenes among their respective isomers. The increasing isomer density for larger values of $x$ is equivalent to the probability of the existence of respective fullerenes.

Together, the above two conditions imply that the probability of formation of closed cage fullerenes with $x$ C atoms $P(C_x) \propto \exp(-\beta E_x + l_0 x)$ where $\beta = 1/kT$. Therefore, the larger the cluster, the higher the probability that it will be a closed cage with increasing isomers. The net effect of $\exp(-\beta E_x + l_0 x)$ on the condensing C vapor is higher proportion of the large fullerenes. The other essential ingredient is $C_2$ that is the agent of transformation; when it is added to a cage, the cage grows to be the next higher cage; if it is excluded then a one-down cage is formed. The time-of-flight spectra from ablated graphite, see for example ref.[1-3], have the higher fullerenes peaked around 100 -150 C, with the lower end around 30-32 C and the maximum determined by the detection capabilities of the mass analyzer. Clearly, an entire range of cages is formed followed by the cycle of growth and shrinkage under right conditions and the environment leading to the Buckyball. Therefore, the higher probability for large $C_x$ (x>60) cages is inherent to the cage formation mechanisms, however, unstable the resulting cage may be. In fact, we show in this analysis that the formation of the large, unstable cages is the basic premise on which a self organizing process for each of the fragmenting cage depends.

**DYNAMICS OF THE TOP-DOWN CAGE SHRINKAGE**

The foregoing dynamical description of the growing fullerenes of ever increasing sizes and higher isomer densities raises two questions: (1) how and why this exponentially growing cage population of ever increasing C-content and diameters stops growing? (2) The above perceived process of growth of cages via $C_2$ addition requires a proficient source of $C_2$s; where do these $C_2$s come from? The proposed model seeks to provide the answers. Especially, one needs an explanation for the downward turning point of the bottom-up growth. As discussed earlier, the number of isomers $I_x$ for each fullerene is seen to increase exponentially. Nonlinear surface forces associated with the local and global curvature of large fullerenes have earlier been shown[14] to induce instability and buckling in the hollow, nano-spheroidal structures. We show that the bottom-up growth of fullerenes with ever increasing C-content, diameters and pentagon-related local deformations is halted by these nonlinear surface forces that initiate cage fragmentation sequences.

A nano-elastic model was developed for the σ-bonded, nano C cages based on the dominance of the stretching effect over bending in the elastic response of C nano shells that creates the pentagonal protrusions[13]. It was shown that the deformations in nano spheroidal shells create internal stresses that have a direct bearing on their structural stability. Curvature-related properties were obtained for the stability of a range of fullerenes from $C_{60}$ to $C_{1500}$. The critical stress $P_{crit}$ was evaluated beyond which the structure becomes unstable. The critical stress has nonlinear dependence on the deformation parameter. This is a crucial feature of large fullerenes that is a direct result of the spherical curvature. Fullerenes-the graphene shells of constant thickness $t$ but with variable radii $R_x$ have nonlinear forces associated with curvature ($1/R_x$) of the surface protrusions. The nano-elastic model[14] of fullerenes with pentagonal protrusions superimposed on a spheroidal structure presents two types of curvatures; the first dealing with the local, pentagon region and the second with the global spheroid. Curvature being inversely proportional to the radius has larger strain- induced surface forces around the pentagons shown in Fig. 3. These local stresses generate instability. The critical stress is estimated as $P_{crit} \propto \zeta^{-1/2} R_x^{-2}$, where $\zeta$ is a measure of the protrusion, $R_x$–the radius of the respective fullerene. $P_{crit}$ has a nonlinear relationship with $\zeta$ yielding an unstable equilibrium. This provides us with the inbuilt structural forces that limit the cage size by initiating the shrinking process. The earlier process of the ever increasing, C-accreting cages is checked and halted by the surface deformation related forces. The bottom-up grown structure may enter into the regime of instability. The top-down trimming starts for the over-grown, pentagonal protrusion ridden closed cages by the emission of $C_2$ via the cage shrinkage route $C_x \rightarrow C_{x-2} + C_2$. That answers the first question asked above. The three well known fullerenes $C_{20}$, $C_{60}$, $C_{540}$ have the same icosahedral symmetry $I_h$ yet only one isomer of each belongs to $I_h$: in the case of $C_{20}$, 1 out of 1 isomer; for $C_{60}$, 1 out of 1812 isomers and 1 out of ~$3 \times 10^{23}$ for $C_{540}$! These have very different surface topographies with only $C_{60}$ having uniform distribution of the strain due to curvature of the perfect spherical shell. Surface deformations of all others lead to non-uniformity of the structural strain. High symmetry is one of the fundamental conditions with greater probabilities of the spheroidal cages' survival in hot carbon vapor but, it is not the only criterion as we discussed in detail elsewhere[14,16].

## THE C$_2$ GAS

The surface deformation induced strain is partially released with the emission of a C$_2$ per fragmentation step $f$ via $C_x \rightarrow C_{x-2} + C_2$. The density of C$_2$ increases as a function of $f$ until each cluster has been transformed into a C$_{60}$ and $(x/2 - 30)$ C$_2$ molecules through a successive, cumulative fragmentation sequences $C_x \rightarrow C_{60} + (x/2 - 30)C_2$. The dynamics of the top-down fragmentations is mapped by the grand canonical ensemble of large cages and the associated C$_2$ gas. From the partition function of the grand canonical ensemble of fragmenting fullerenes, entropy is calculated at each fragmentation stage which is shown to be 2 to 3 orders of magnitude less than the entropy generated by the associated C$_2$ gas. The initial bottom-up sequence of large cage formation is followed by the top-down cage shrinking cascades. Together, these are shown to constitute the self organizing carbon cages with initial growth followed by surface instability-induced fragmentations. These dynamic processes are shown to start the process of self organization of fullerenes in carbon vapor leading to the Buckyball.

## FULLERENES AS ROTORS

Addition of pentagons induces curvature in flat graphene-like sheets during the initial growth phase. Alternatively, if a pentagon is the starting structure, then a corrannulene bowl with the inherent curvature results; its further growth may yield a cage. Bending moment associated with the curved, bowl-like structure will set it into rotation[14]. Carbon accretion thereafter, takes place into the rotating, curved structure until it closes. After the cage closure, the fullerenes have rotational moment of inertia $M_x = 2mR_x^2/5$, where m is the mass and $R_x$ the cage radius. In the ensemble of rotating fullerenes, the rotational partition function for each of the fullerene with $M_x$ is $z_{rot} = (8\pi^2 kTM_x / \hbar^2)^{3/2} = T/\theta_{rot}$, with rotational temperature $\theta_{rot} = \hbar^2 / 2M_x k$. The associated entropy $S_x = k \ln([z_{rot}]^x / x!)$ [15]. The rotation of the primary ensemble's fullerenes allows calculations of their entropies during the subsequent fragmentation sequences. Sum of entropies of all the fullerenes as a function of the fragmentation steps $\sum S_x(f)$ reveals the dynamical features of the self organizing cages. For all fullerenes larger than $C_{60}$, two mechanisms compete to reduce the structural strain; one occurs via SW transformation[17] for the re-arrangement of C bonds while the other is fragmentation with the emission of a $C_2$. Bonds rearrangement redistributes spheroid's strain and the processes are energy consuming for all isomers. Intuitively, a cage will go through a sequence of steps including both the options. Fragmentation not only reduces the surface area but also leaves the cage with fewer atoms to be rearranged. The process is presumed to continue in the primary ensemble of hot, collision dominated, internal stress- ridden fullerenes. The fragmenting cages are in contact with a hot gas of $C_2$ whose density increases with each fragmentation stage. With $C_2$ linked with every fragmentation step, its increasing density $\propto$ the decrease of the cages' surface area. The hot $C_2$ gas of growing density becomes an integral component of the grand canonical ensemble. The statistical mechanical treatment of the cages and the $C_2$ gas as a function of the fragmentation stages describes the self organization whose outcome is deformation-free, perfect spheroidal $C_{60}$. Mapping of the dynamics of the grand canonical ensemble as function of cage fragmentation reveals that the positive entropy of the ensuing $C_2$ gas compensates for the shrinking cages' negative entropy. The total, positive entropy of the fragmenting fullerenes and $C_2$ gas during fragmentation stages continuously increases. The cumulative entropy of various sets of fullerenes, however, decreases.

**THE SHRINKING CAGES-PULSED MODE**

The present model is based on three assumptions that: (1) the laser ablated graphite produces a pulse of closed cages in the same ratio as that of their respective isomers i.e. $N_x \approx I_x$.

(2) The range of fullerenes considered is between $C_{60}$ and $C_{100}$. (3) At each fragmentation step half of all the cages fragment through $C_x \rightarrow C_{x-2} + C_2$. $C_{60}$ being the only exception and acts as an attractor. Fig. 4(a) shows results from 45 successive fragmentation steps of 3D evolution of the distribution of the isomer-based cage population $N_x(f)$ for respective fullerenes. The original cage distribution in the primary ensemble is at $f = 0$, with each successive $f$ changing the fullerene population and characterizing the dissipative dynamics of the larger ones fragmenting to enrich the next smaller neighbor's population. The process continues with all the larger fullerenes shrinking by spitting out a $C_2$ per step per cage. By $f = 41$, almost all larger cages can be seen shrinking towards the attractor i.e. $C_{60}$. For each larger cage only one $C_{60}$ can be obtained while $(x/2 - 30) C_2$s are emitted. The density $N_{C_2}(f)$ of the accumulating gas of $C_2$ as a function of $f$ shows a monotonous increase. However, the total fullerene density is preserved. The rate of increase of $N_{C_2}(f)$ reduces to zero for $f > 40$ when majority of the cages have been transformed into $C_{60}$. The pattern of change of population densities $N_x(f)$ of the five chosen fullerenes $C_{100}$, $C_{80}$, $C_{70}$, $C_{62}$ and $C_{60}$ shows that all species, except $C_{60}$, go through a maximum and then reduce to zero within 45 fragmentation steps in Fig. 5. An exponentially increasing function $\exp(l_1 f)$ can be fitted to the profile of $N_{60}(f)$ plotted as a function of $f$; $l_1$ being the coefficient of stability for this dynamic process for the exponential increase of the density of $C_{60}$.

The simulation of the dynamics of cluster fragmentation of the primary fullerene ensembles formed in a pulsed ablation experiment can illustrate various sequences and stages of the shrinking and $C_2$-spitting cages. $C_{100}$ population can be seen to reduce significantly within the first 5 steps while that of $C_{80}$ goes through a peak around the 18th step and then diminish by the 30th. The larger fullerenes gradually disappear by enriching their next smaller counterparts. The $C_{62}$ density continues to rise up to $f = 35$ and then reduces and acts as a one-way "gate" for the increasing density of $C_{60}$ which is the sole survivor and attractor of the dynamical system. After about 50 steps the final $C_{60}$ population ≈ the sum of all the isomers of the primary fullerene ensemble $\sum I_x$. In our assumption of $60 \leq x \leq 100$ yielding $\sum I_x \cong 1.45 x 10^5$, therefore, that is the maximum possible number of $C_{60}$ per ablation.

The variations of the rotational entropies $S_x(f)$ of all isomers of five selected fullerenes provide insight into the details of the fragmentation dynamics with $S_{100}(f)$ reducing to zero within the first 10 fragmentation steps while that of $C_{80}$, $C_{70}$, $C_{62}$ and $C_{60}$ first increase to a peak and then reduce as shown in Fig. 6(a). $S_{60}(f)$ becomes negative for $f \geq 30$. The sum of entropies of all the fullerenes $\sum S_x(f)$ from $C_{60}$ to $C_{100}$, at each fragmentation stage for the set of 21 fullerenes i.e. $C_{100}$, $C_{98}$, $C_{96}$....$C_{64}$, $C_{62}$, $C_{60}$, shows a net decrease of $\sum S_x(f)$ with the increasing $f$ indicating that the entropy reduction is the hallmark of the inter-fullerene transformation plotted in Fig. 6(a). The cumulative entropy $S_{C_2}(f)$ of the growing number of $C_2$s is positive and increasing in Fig. 6(b). $S_{tot}(f) = \sum S_x(f) + S_{C_2}(f)$ being the total entropy of the grand canonical ensemble comprising of all the fullerenes and the associated $C_2$ gas is also positive and increases by each fragmentation step. Calculations were performed at 1000 K and all entropies are in units of k-the Boltzmann constant.

## THE CONTINUOUS ARC-DISCHARGE MODE

The model developed here for the pulsed laser ablation can be extended to the continuous vapor created in the typical arc discharge. In that case, $C_{100}$ becomes the source and its population does not decrease at each fragmentation stage. In this case the energy deposited in the arc discharge creates the continuous C vapor that is condensing into the cages all the time. Here one has to describe the simultaneous formation, fragmentation and re-formation of the closed cages as a continuous process. In the pulsed case scenario discussed above, condensation results in the formation of the fullerene isomer distribution that subsequently goes through the cage shrinking sequences of this primary ensemble. Whereas, in the continuous arc discharge, the constancy of the source density i.e. $C_{100}$, is maintained. $C_{100}$ acts as the source. The cage fragmentation is a serial process i.e., $C_{100} \rightarrow C_{98} \rightarrow C_{96} ...... C_{62} \rightarrow C_{60}$. After the initial fragmentation stages the density of $C_{62}$ remains constant and $\approx \sum I_x$ where x is between 62 and 100. Same is the case for the higher cages. Figure 7 shows the entropy directly related with the increasing density of $C_2$ as a function of $f$. $N_x(f)$ for $x = 100, 80, 70, 62$ and 60, against $f$ show similar pattern as seen in figure 5. The inset in Fig. 7(a) shows the calculated entropy of the each fullerene $S_x(f)$ as a function of $f$. Unlike the pulsed case in Fig. 6(a), the entropies $S_{100}(f), S_{80}(f), S_{70}(f)$ and $S_{62}(f)$ increase with $f$. In the continuous discharge case, $S_{60}(f)$ initially increases, as in figure 6(a), and then decreases beyond $f \geq 20$ eventually becoming negative. This behavior of $C_{60}$ affects the sum of the entropies of all fullerenes $\sum_{60}^{100} S_x(f)$ shown in Fig. 7(a). The sum of entropies of all fullerene cages increase up to $f \geq 20$, has a broad plateau and then decreases. The associated entropy of the cumulative gas of $C_2$, $S_{C_2}$ shows an increasing trend that is about three orders of magnitude higher than the sum of the entropies of all the rotating cages, is shown in the inset of Fig. 7(b). The total entropy $S_{tot}(f)$ of all the rotating, shrinking cages and the $C_2$ liberated in the fragmentation process shows a monotonous increase in figure 7.

**CONCLUSIONS**

The dynamics of cluster growth in the condensing carbon vapor is shown to be dominated by large closed cages whose respective heats of formation reduce with the increasing carbon content while the number of isomers for each fullerene increases exponentially as a function of *x*. We have shown that the bottom-up dynamics of condensing carbon vapor into a wide range of ensembles of fullerenes is not sensitive to the initial starting conditions. We identify a dynamical system that has built-in chaotic conditions for each growth step or the fragmentation stage, but the net outcome of the overall dynamics is predictable and well defined. Take the example of three successive cages $C_{88}$, $C_{90}$ and $C_{92}$ with respective isomers number densities as 81, 738, 99,918 and 126, 409[11]. When any of the 99,918 isomers of $C_{90}$ absorbs a $C_2$, it has 126, 409 isomer shapes available to choose from. Any subsequent absorption will also have the same number of possibilities, yet no two outcomes are likely to be exactly the same. Same is true for the higher cages, where this probability decreases further. This can be defined as the horizontal dimension of our dynamical system and is chaotic[18]. The horizontal chaos increases with the cage size on each absorption of a $C_2$. While in the case of a larger cage fragmenting by spitting out a $C_2$, it has a reducing number of isomers available at each fragmentation stage. If any isomer out of the 99,918 $C_{90}$s fragments, it has a fewer (i.e. 81,738) options available to become a $C_{88}$. The numbers of options reduce as the cage loses further $C_2$s. Further cage fragmentation, in the top down sequence, with reducing shape options, will pile up the isomers with similar $x$. Similarly, $C_{70} \rightarrow C_{68} + C_2$ reduces the total shape options from 8,149 to 6,332. This is the cage predictability due to the variations in $x$; defined as the vertical dimension that defines the dynamics of cage shrinkage. Self organization against the $x$ parameter-the vertical dimension, is well defined and predictable. On the other hand, the horizontal movement along $I_x$-the isomeric densities has a built in chaotic character.

Nonlinear surface forces associated with the local and global curvature of large fullerenes induce instability and buckling halting their carbon accretion and growth. During the ensuing cage fragmentation sequences the emission of a $C_2$ per step relieves the local strain. $C_2$ emissions shrink large cages eventually into $C_{60}$. Top-down sequences of fragmentations via shrinking of large cages of the grand canonical ensemble result in the increasing density of the associated $C_2$ gas. At each fragmentation step the ensembles' entropy was mapped for all the shrinking fullerenes along with the positive entropy generated by the $C_2$s. The initial bottom-up sequence of large cage formation is shown to be followed by the subsequent top-down shrinking cascades. Together, these are shown to constitute the self organizing, large, carbon cages. The initial C-accreting stage and subsequently for the accumulation of $C_{60}$s indicate stable dynamics of the self organizing fullerenes into $C_{60}$. The mechanism for conversion of the intra-fullerene bonds that lead non-icosahedral ones to the icosahedral $C_{60}$ where out of a total of 1812 isomers has already been extensively debated in the fullerene literature[7]. We conclude that by creating a hot, fullerene forming and fragmenting environment where for each cage a large number of $C_2$ molecules actively participate by absorption and emission. Such a grand canonical ensemble has been treated by assuming the cages to be 3D rotors along with the evolving and an all-encompassing $C_2$ gas. It is a unique combination of a continuous ongoing phase transformation that has a well-defined output. The driving force for these transformations is the curvature-related local surface deformations of the cages. Such a dynamical description of self organization may have important implications for understanding similar processes elsewhere in nature.

**Figure Captions**

**FIG. 1.** The heat of formation per C atom, $E_x$ is plotted for fullerenes as a function of the number of C atoms $x$. The cages for $x=24$ to 60 are shown for three possible routes of fullerene formation[13]. The inset shows the trend up to $x=100$.

**FIG. 2.** Fullerene isomers $I_x$ increase exponentially with $x$ ($\propto \exp(l_0 x)$); $l_0$ is an exponent associated with the dynamics of the multiplicity of fullerene structures with the same number of C atoms i.e., isomers.

**FIG. 3.** Two cage stability criteria are shown as a function of $x$ using the continuum[14] and quantum mechanical models[16]. The critical stress $P_{crit} \propto \zeta^{-1/2} R_x^{-2}$, has a nonlinear dependence on $\zeta$. Alternatively, the delocalized π electrons of fullerenes can be treated as a Fermi gas whose degeneracy pressure $P_\pi$ is a measure of cage stability. **Inset:** A closed cage with radius $R_x$, thickness of σ- shell $t$ and $\zeta$ -the protrusion; the fullerene is assumed to be a 3D rotor

**FIG. 4.** (a) 3D plot of the dynamics of cage shrinkage via $C_x \rightarrow C_{x-2} + C_2$ shows self organization of large C cages with half of all cages fragmenting at each step. The starting cage distribution in the primary grand canonical ensemble is at $f = 0$. At each successive $f$ all larger fullerenes shrink by spitting out a $C_2$ per step. (b) 2D plot of the shrinking sequences as shown in (a). Three fragmentation steps $f=0$, 21 and 41 illustrate the top-down sequence. The box around $C_{60}$ shows a delta function.

**FIG. 5.** Number densities of $C_{100}$, $C_{80}$, $C_{70}$, $C_{62}$ go through a maximum and then reduce to zero within 45 fragmentation steps. The exponential increase of $C_{60}$ density as a function of $f$ is $N_{60}(f) \sim \exp(l_1 f)$; $l_1$ being a stability index for this dynamic sequence.

**FIG. 6.** (a) The sum of entropies of all the fullerenes $\sum S_x(f)$ from $C_{60}$ to $C_{100}$ shows a net decrease with increasing $f$. The inset shows variation of $S_x(f)$ for the five selected fullerenes $C_{100}$, $C_{80}$, $C_{70}$, $C_{62}$ and $C_{60}$. $S_{60}(f)$ is negative for $f \geq 30$. (b) $S_{C_2}(f)$ of the $C_2$s is positive and increasing as shown in the inset while the total entropy $S_{tot}(f) = \sum S_x(f) + S_{C_2}(f)$ of fullerenes and $C_2$ gas is also positive and increases with $f$.

**FIG. 7.** (a) Inset: $S_x(f)$ of five selected fullerenes increase to become constant while that of $C_{60}$ first increase to a peak around $f \geq 20$ and it becomes negative for $f \geq 28$. The trend of sum of entropies of all the fullerenes $\sum S_x(f)$ from $C_{60}$ to $C_{100}$ shows a gradual increase up to $f \approx 20$ followed by a plateau and then a net decrease of entropy with the increasing $f$. **(b)** Inset: $S_{C_2}(f)$ of the $C_2$s is positive and increasing. The total entropy $S_{tot}(f) = \sum S_x(f) + S_{C_2}(f)$ of the grand canonical ensemble comprising of all the fullerenes and the associated $C_2$ gas shown is also positive and increases with $f$.

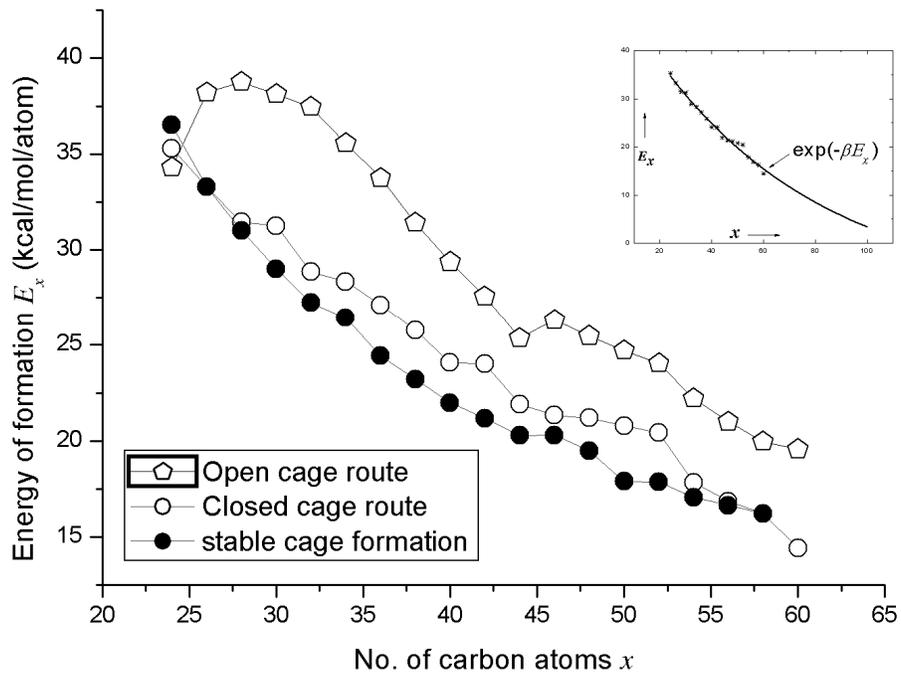

FIG. 1.

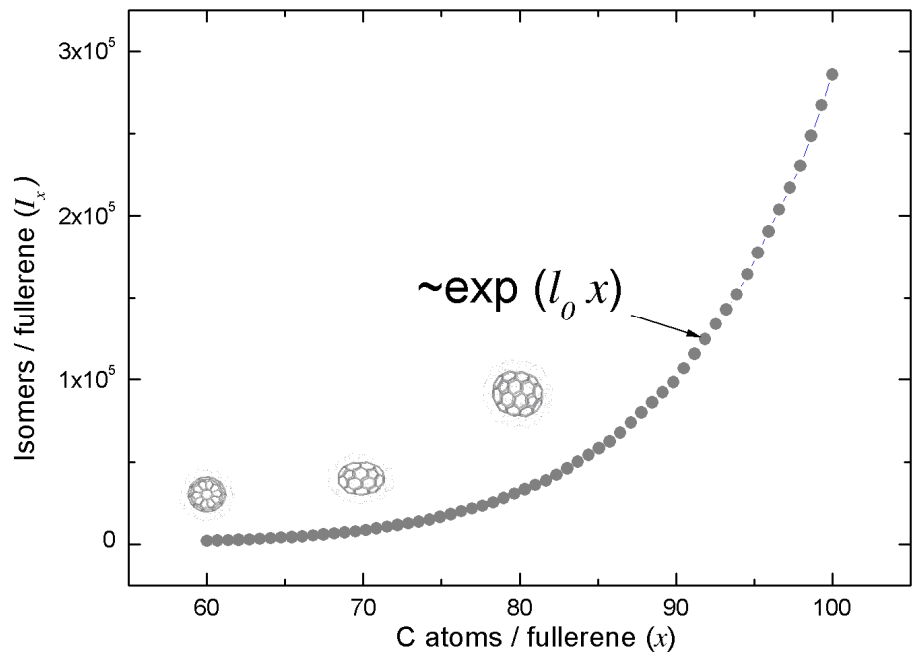

FIG. 2.

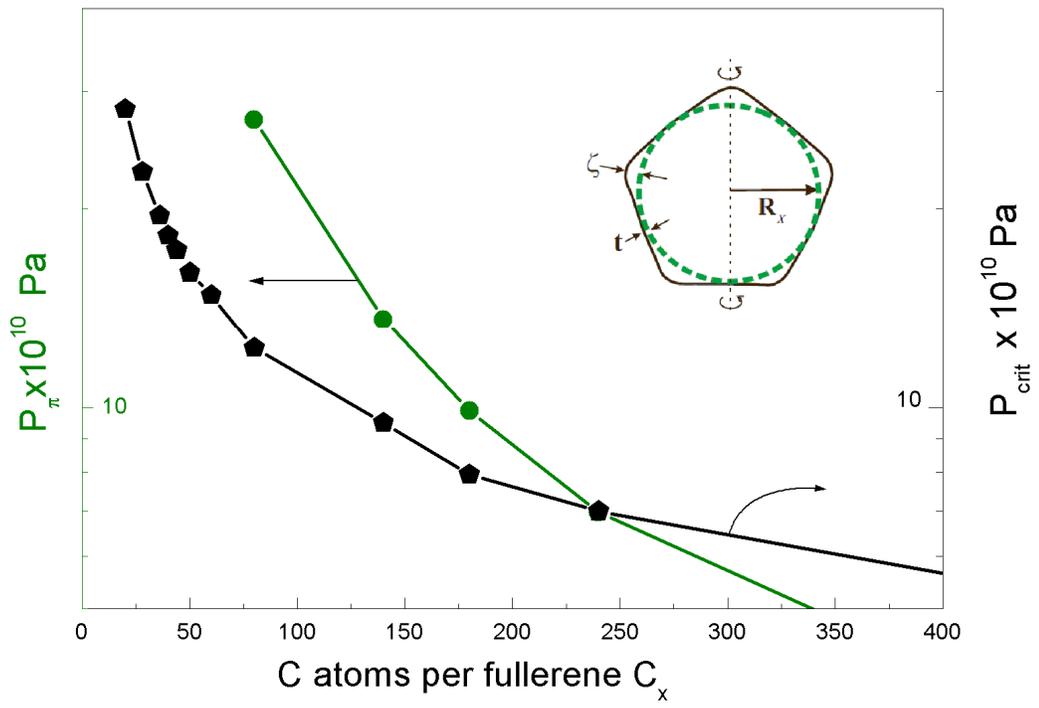

FIG. 3.

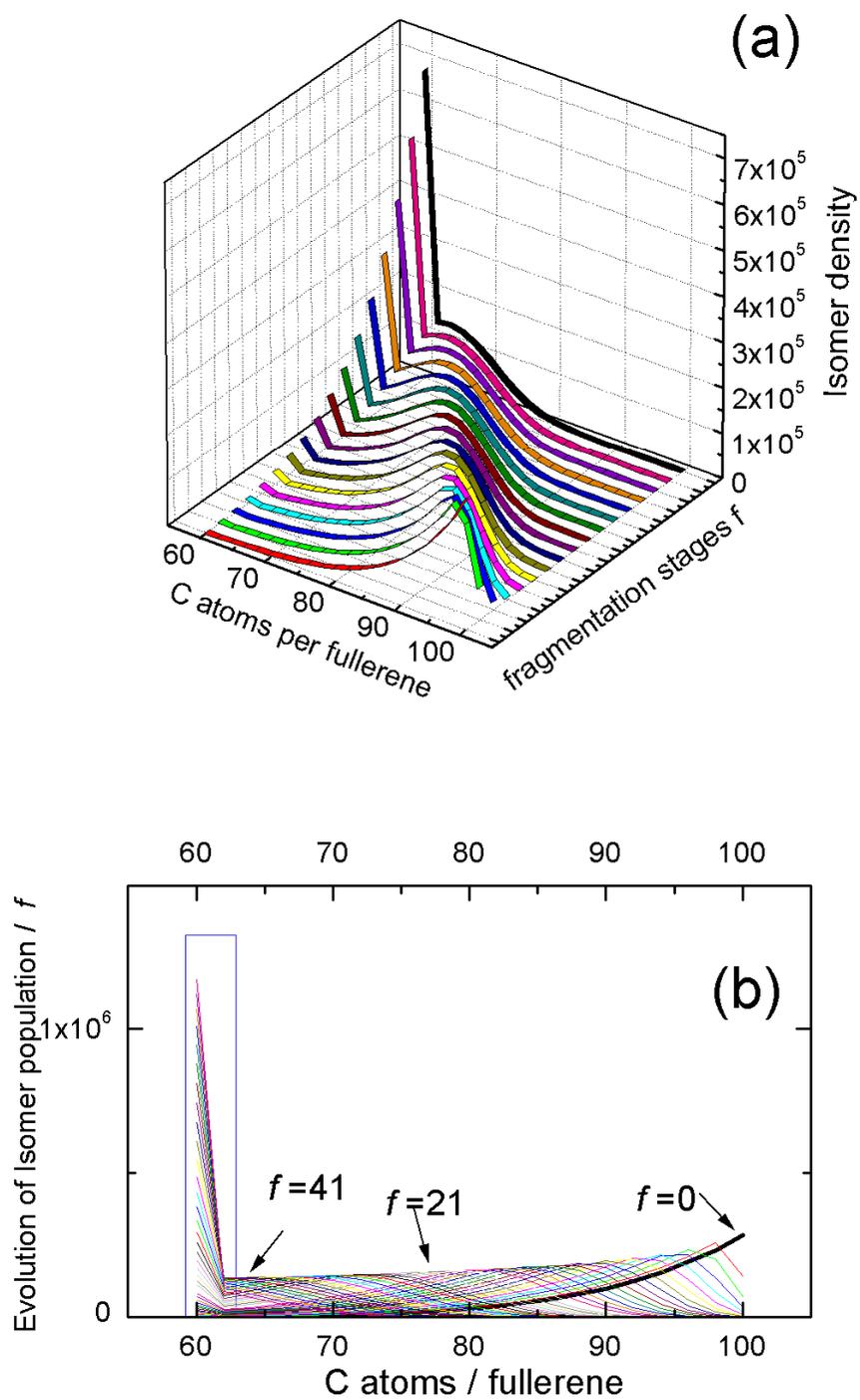

FIG. 4.

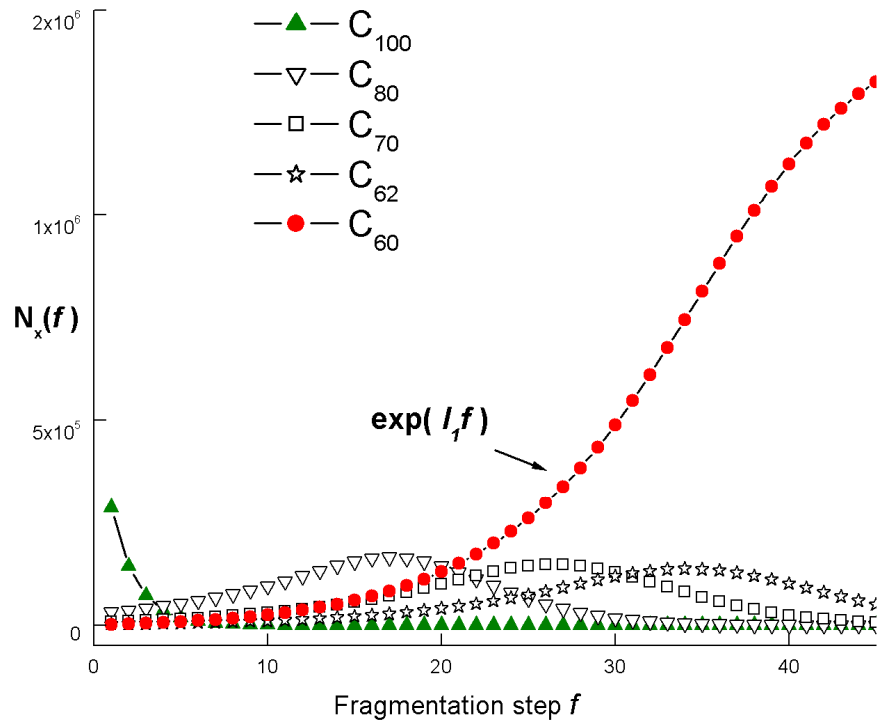

FIG. 5.

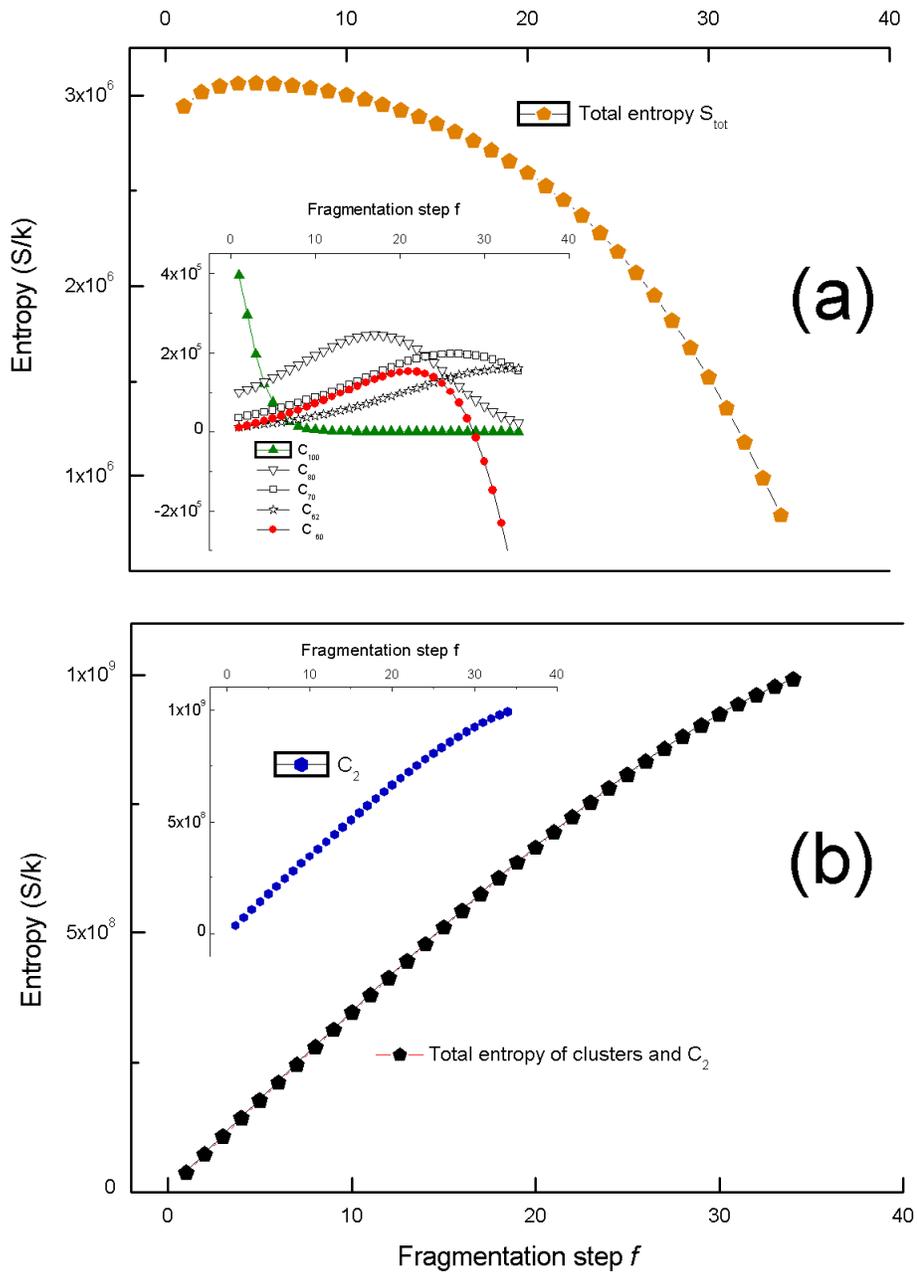

FIG. 6.

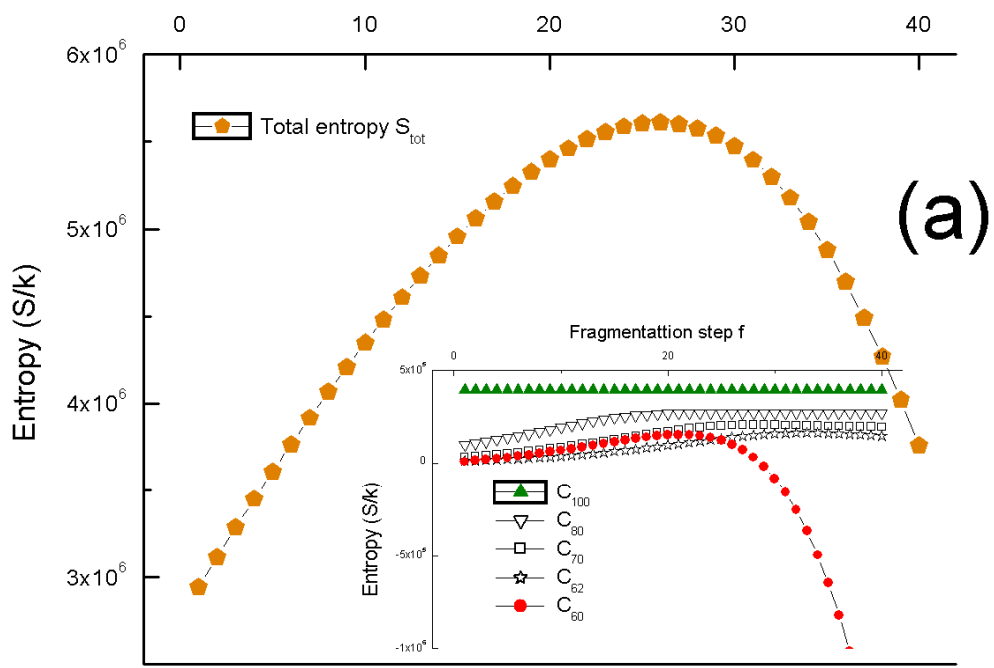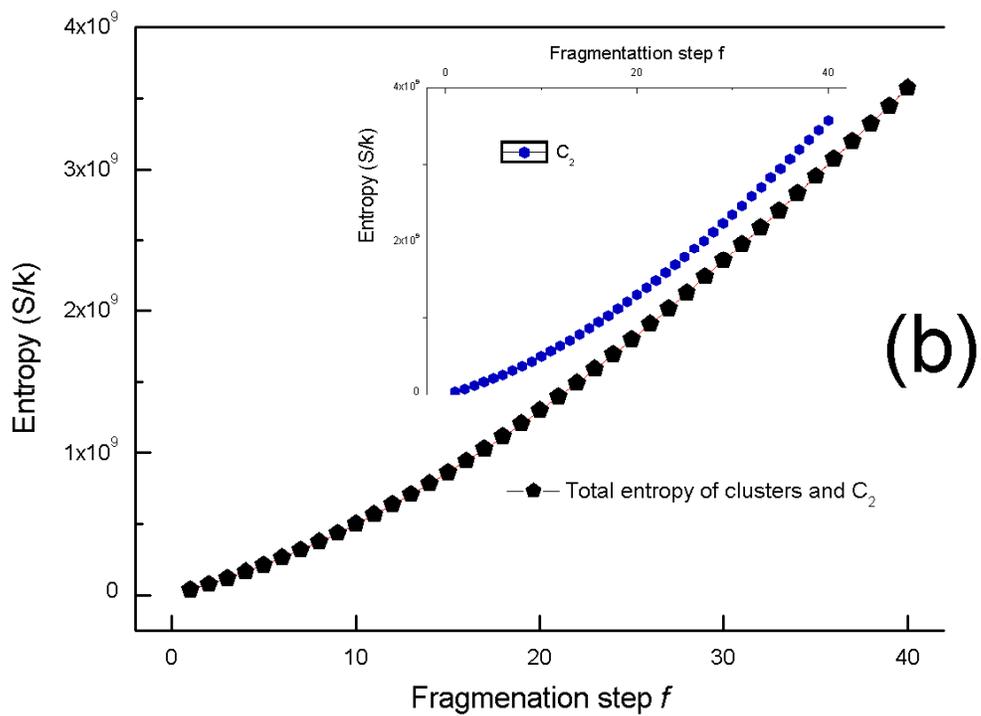

FIG. 7.